\documentclass[showpacs,twocolumn]{revtex4}

\usepackage{graphicx}
\usepackage{dcolumn}
\usepackage{amsmath}

\makeatletter
\def\btt#1{\texttt{\@backslashchar#1}}
\DeclareRobustCommand\bblash{\btt{\@backslashchar}} \makeatother

\begin{document}

\title[]{Radiating Kerr-like regular black hole}
\author{Sushant~G.~Ghosh$^{a,\;b\;}$} \email{sghosh2@jmi.ac.in,
sgghosh@gmail.com}
\author{Sunil~D.~Maharaj$^{a}$}\email{maharaj@ukzn.ac.za} \affiliation{$^{a}$ Astrophysics and Cosmology
Research Unit, School of Mathematical Sciences, University of
KwaZulu-Natal, Private Bag 54001, Durban 4000, South Africa}
\affiliation{$^{b}$ Center for Theoretical Physics, Jamia Millia
Islamia, New Delhi 110025, India}
\date{\today}

\begin{abstract}
We derive a radiating regular rotating black hole solution,
radiating Kerr-like regular black hole solution. We achieve this by
starting from the Hayward regular black hole solution via a complex
transformation suggested by Newman-Janis. The  radiating Kerr
metric, the Kerr-like regular black hole and the standard Kerr
metric are regained in the appropriate limits. The structure of the
horizon-like surfaces are also determined.
\end{abstract}



\maketitle

\section{Introduction}
The formation of spacetime  singularities is a quite common
phenomenon in general relativity and, indeed, celebrated theorems,
proved by Penrose and Hawking \cite{he}, state under some
circumstances singularities are inevitable in general relativity. As
these theorems uses only the laws of general relativity and some
properties of matter, they are valid generally. It is widely
accepted that spacetime singularities do not exist in Nature; they
 are limitation or creation of the classical theory. The
existence of a singularity implies there exists a point in spacetime
where the laws of physics break down or signal a failure of the
physical laws. It turns out that what amount to a singularity in
general relativity could be adequately explained by some other
theory. If physical laws do exist at those extreme situations, then
we should route to a theory of quantum gravity. However, we are yet
distant away from a definite theory of quantum gravity. So a line of
action is to understand the inside of a black hole and resolve its
singularity by carrying out research of classical or semi-classical
black holes, with regular, i.e., nonsingular, properties. This can
be motivated by quantum arguments.   Sakharov \cite{ads} and Gliner
\cite{Gliner} proposed that spacetime in the highly dense central
region of a black hole should be de Sitter-like for $r \simeq 0$
(see also, \cite{Mukhanov}). This indicates that an unlimited
increase of spacetime curvature during a collapse process can lead
to the stop of the collapse if quantum fluctuations dominate the
process. This places an upper bound on the value of the curvature
and obliges the formation of a central core.

Bardeen \cite{Bardeen} realized concretely  the idea of a central
matter core, by proposing the first regular black hole solution of
the Einstein equations.  Bardeen's regular metric is a solution of
the Einstein equations in the presence of an electromagnetic field,
yielding a alteration of the Reissner-Nordstr$\ddot{o}$m metric. But
near the center the solution tended to a de Sitter core solution.
Subsequently, there has been enormous development in investigating
the properties of regular black hole solutions
\cite{sa,AGB,Hayward,regular}, but most of these regular black hole
solutions are more or less based on Bardeen's proposal. In
particular, an interesting proposal is made by Hayward
\cite{Hayward} for the formation and evaporation of regular black
holes, in which the static region is the Bardeen-like black hole.
The dynamic regions are Vaidya-like black hole, with negative energy
flux during evaporation and ingoing radiation of positive energy
flux during collapse. The latter is balanced by outgoing radiation
of positive energy flux and a surface pressure at a pair creation
surface.  This is the only non-stationary or dynamical regular black
hole.   However, these non-rotating metrics can not be  tested by
astrophysical observations, as the black hole spin plays an
important and fundamental role in any astrophysical process.

The generalization of these stationary regular black holes to the
axially symmetric case, the Kerr-like regular black hole, was
addressed recently \cite{Bambi,Neves:2014aba,Toshmatov:2014nya}. In
particular, it was established \cite{Bambi,Toshmatov:2014nya} that
the rotating regular black hole solutions can be obtained starting
from regular black hole solutions by a complex coordinate
transformation previously suggested by Newman and Janis \cite{nja}.
However, this is obviously not the most physical scenario and we
would like to consider dynamical black hole solutions, i.e., black
holes with non-trivial time dependence. Further, the axially
symmetric counterpart of the regular Vaidya-like black hole is still
unexplored, e.g., the radiating generalization of the regular
Kerr-like black hole is still unknown. It is the purpose of this
paper to obtain this metric. Thus we extend a recent work
\cite{Hayward} on radiating regular black holes to include rotation,
and it is also a non-static generalization of the Kerr-like regular
black hole solution \cite{Bambi}.  We also carry out detailed
analysis of the horizon structure of radiating Kerr-like regular
black holes, which also valid for static Kerr-like regular black
hole and not done earlier. It should be pointed out that the Kerr
metric \cite{kerr} is undoubtedly the single most significant exact
solution in the Einstein theory of general relativity, which
represents the prototypical black hole that can arise from
gravitational collapse. The radiating or non-static counterpart of
the Kerr black hole was obtained by Carmeli \cite{Carmeli}.  We also
show that  the Kerr-like regular black hole, the Kerr black hole and
the radiating  Kerr-like black hole arise as special cases of the
radiating Kerr-like regular black hole.

In this paper, we obtain a  radiating
Kerr-like regular metric  in Section II.
The Newman-Janis algorithm is applied to spherically symmetric
radiating solutions, and radiating rotating solutions are obtained. We  investigate the structure
and locations of horizons of the radiating Kerr-like regular metric
in Section III.
 The paper
ends with concluding remarks in Section IV.   We use units which fix the speed of light and the
gravitational constant via $G = c = 1$, and use the metric signature
($+,\;-,\;-,\;-$).

\section{Radiating Rotating black hole via Newman-Janis}
We wish to obtain a radiating rotating regular black hole
solution from spherically symmetric black hole solutions via the complex
transformation suggested by Newman-Janis \cite{nja}. For this
purpose, we begin with  the "seed metric", expressed in terms of the
Eddington (ingoing) coordinate $v$, as
\begin{equation}
ds^{2} = e^{\psi(v,r)}dv\left[f(v,r)e^{\psi(v,r)}dv + 2  dr\right] -
r^2 d\Omega^2, \label{Metric}
\end{equation}
with $d\Omega^2 = d \theta^2+ \sin^2 \theta d \phi^2$. Here
$e^{\psi(v,r)}$ is an arbitrary function. It is useful to introduce
a local mass function $m(v,r)$ defined by $f(v,r) = 1 - {2
m(v,r)}/{r}$. For $m(v,r) = M(v)$ and $\psi(v,r)=0$, the metric
reduces to the standard Vaidya metric. We can always set without any
loss of generality, $\psi(v,r) = 0.$  Thus any spherically symmetric
radiating black hole is defined by the metric (\ref{Metric}).  The function
$f(v,r)$ is function of $v$ and $r$, and depends on the matter field.

The Newman-Janis algorithm can be applied to any spherically
symmetric static black hole solution of general relativity to
generate rotating black hole spacetimes. For example, the Kerr
metric can be obtained from the Schwarzschild metric, and the
Reissner-Nordstr$\ddot{o}$m solution leads to the Kerr-Newman
solutions, which is based on a complex coordinate transformation.
Recently, regular rotating black holes were derived from exact
spherically symmetric regular black hole solutions \cite{Bambi}.
 In what follows, we extend to apply the Newman-Janis
algorithm to the general spherically symmetric radiating
$\textit{seed}$ metric (\ref{Metric}) which can be put in the form
\begin{equation}
ds^{2} = f(v,r) dv^2 + 2  dv dr -
r^2 d\Omega^2, \label{Metric1}
\end{equation}
to construct a general radiating rotating black hole solution. First
step of the Newman-Janis algorithm is not required here as the
$\textit{seed}$ metric (\ref{Metric1}) is already in the
Eddington-Finkelstein coordinates.

The metric $\tilde{g}_{ab}$ given by Eq. (\ref{Metric}) can be
written in terms of a null tetrad \cite{nja} as
\begin{equation}
\tilde{g}^{ab} = -L^a N^b - L^b N^a + M^a \bar{M}^b + M^b \bar{M}^a,
\label{NPmetric}
\end{equation}
where the null tetrad has the form \begin{eqnarray*}
L^a &=& \delta^a_r,\\
N^a &=&  \delta^a_u - \frac{1}{2} f(v,r) \delta^a_r ,\\
M^a &=& \frac{1}{\sqrt{2}r} \left( \delta^a_{\theta}
  + \frac{i}{\sin\theta} \delta^a_{\phi} \right).
\end{eqnarray*}
This tetrad is orthonormal obeying the conditions \begin{eqnarray}
&&L_a M^a = L_a \bar{M}^a = N_a M^a = N_a \bar{M}^a = 0, \\
&&L_a L^a = N_a N^a = M_a M^a = \bar{M}_a \bar{M}^a = 0, \\
&&L_a N^a = -1,~~~ M_a \bar{M}^a = 1. \end{eqnarray}
Now we allow for some $r$ factors in the null vectors to take on complex values. We rewrite the null vectors in the form \cite{nja,d'Inverno,sgsd}
\begin{eqnarray*}\label{NPjnw}
L^a &=& \delta^a_r, \\
N^a &=& \left[ \delta^a_u - \frac{1}{2}f(v,r,\bar{r}) \delta^a_r \right], \\
M^a &=& \frac{1}{\sqrt{2} \bar{r}}   \left( \delta^a_{\theta}
  + \frac{i}{\sin\theta} \delta^a_{\phi} \right).
\end{eqnarray*}
 Following the Newman-Janis prescription
\cite{nja}, we now write
\begin{equation}\label{transf}
{x'}^{\mu} = x^{\mu} + ia (\Delta^{\mu} - \delta_u^{\mu})
\cos\theta \rightarrow \\ \left\{\begin{array}{ll}
v' = v - ia\cos\theta, \\
r' = r + ia\cos\theta, \\
\theta' = \theta,~~~\phi' = \phi. \end{array}\right.
\end{equation}
we also transform the tetrad $Z^a_s = (L^a,\; N^a,\;M^a,\; \bar{M}^a)$ in the usual way
\begin{equation} Z'^a_s =
\frac{\partial x'^a}{\partial x^b} Z^b_s,
\end{equation} leading to
\begin{eqnarray}
L^a &=& \delta^a_r, \\
N^a &=&   \left[ \delta^a_v - \frac{1}{2} \mathcal{F}(v,r,\theta) \delta^a_r \right], \\
M^a &=& \frac{1}{\sqrt{2}(r+ia\cos\theta)} \left[ ia\sin\theta \left(\delta^a_v - \delta^a_r \right) + \delta^a_\theta + \frac{i}{\sin\theta} \delta^a_\phi \right], \end{eqnarray}
 and, dropping the primes.
 This transformed tetrad yields a new metric (see Ref. \cite{nja,sgsd}, for further details) given by the line
 element
\begin{eqnarray}\label{rotbh}
ds^2 & = & \mathcal{F}(v,r,\theta) dv^2 + 2 dv dr - \Sigma(r,\theta) d
\theta^2 - 2 a \sin^2 \theta dr d\phi  \nonumber \\ & + & \left[ a^2
( \mathcal{ F}(v,r,\theta) -2 )\sin^2 \theta  - \Sigma(r,\theta)
\right] \sin^2 \theta d\phi^2 \nonumber \\ & + & 2 a \left[ 1 -
\mathcal{ F}(v,r,\theta) \right] \sin^2 \theta dv \; d\phi.
\end{eqnarray}
Here $\mathcal{F}(v,r,\theta) $ is function which depends on
$f(r,v)$.  It describes the exterior field of the radiating rotating
objects.  We have applied the aforesaid procedure to radiating
models. But, the method is general and is applicable to any general
radiating spherically symmetric solution to generate a general
rotating radiating spacetimes (\ref{rotbh}). Carmeli \cite{Carmeli}
was first obtained the Metrics of rotating radiating spacetime,
which in the limit $a=0$ reduces to Vaidya spacetime.  To further
support our analysis, we should be able to rediscover the solution
obtained by Carmeli \cite{Carmeli}, but by using Newman-Janis
algorithm. In the Vaidya case
\begin{equation}\label{fvr}
f(v,r) = 1 - \frac{2M(v)}{r}.
\end{equation} After complex transformations it has the form $$\mathcal{F}(v,r,\theta) =
1 - \frac{2M(v)r }{\Sigma },$$  and $\Sigma = r^2 + a^2
\cos^2\theta$.  In the above analysis, all the steps of Newman-Janis
algorithm are applicable to radiating spacetime to generate the
corresponding  radiating rotating spacetime. However, to generate
the Carmeli's radiating rotating spacetime, we must demand that the
mass term $M(v)$ remains invariant under the complex
transformations. Then, we start with the radiating spherically
symmetric metric (\ref{Metric}), written in Eddington-Finkelstein
coordinates and performing the Newman-Janis algorithm with the above
$ f(v,r) $ given by (\ref{fvr}), we derive a radiating rotating
solution which takes the form
\begin{eqnarray} \label{rknm}
ds^2&=& \frac{1}{\Sigma}\left[\Delta -  a^2 \sin^2 \theta \right] dv^2+ 2  \left[ dv - a \sin^2 d\phi \right] dr  \nonumber  \\
   & & - {\Sigma} d \theta^2 + \frac{2 a}{\Sigma}\left[\Delta (r^2 + a^2)-1 \right]\sin^2\theta dv d\phi \nonumber \\
   & & - \frac{1}{\Sigma}\left[ (r^2 + a^2)^2 - \Delta
    a^2 \sin^2\theta \right]
   \sin^2\theta
   d\phi^2,
\end{eqnarray}
where
\begin{eqnarray*}
\Sigma &= & r^2+a^2\cos^2\theta,\\
\Delta &=&r^2+a^2-2M(v) r.\\
\end{eqnarray*}
Here $M(v)$ is a function of the retarded time $v$ identified as the
mass of the black hole, and $a$ is the angular momentum per unit
mass.  Thus, the metric (\ref{rknm}) bears the same relation to Kerr
as does the Vaidya metric to the Schwarzschild metric. The metric
(\ref{rknm}) was originally obtained by Carmeli \cite{Carmeli}. Thus
we have a kind of radiating rotating metric or radiating Kerr-like
solution, Hence for definiteness we shall call the metric
(\ref{rknm}) as the radiating Kerr black hole.

In order to further discuss the physical nature of the radiating
Kerr-like black hole, we introduce their kinematical parameters. Following
\cite{bc,jy,rm,bdk,xd98,xd99}, the null-tetrad of the metric
(\ref{rknm}) is of the form
\begin{eqnarray*} \label{nvector}
  l_a &=& \left[ 1,\;0,\;0,\; -  a \sin^2 \theta\right], \\
  n_a &=& \left[1 \frac{\Delta}{2 \Sigma},\;1,\;0,\;  \frac{\Delta}{2 \Sigma}a \sin^2 \theta \right],\\
  m_a &=& \frac{1}{\sqrt{2}\rho}\left[ i a \sin \theta,\; 0,\; \frac{\Sigma}{\Theta},\; - i(r^2+ a^2) \sin \theta  \right],\\
  \bar{m}_a &=& \frac{1}{\sqrt{2}\bar{\rho}}\left[- i a \sin \theta,\; 0,\; \frac{\Sigma}{\Theta},\;  i(r^2+ a^2) \sin \theta
  \right],
\end{eqnarray*}
where $\rho = r + i a \cos \theta $ and $\bar{\rho}$  is it's complex conjugate. The null tetrad obeys null,
orthogonal and metric conditions
\begin{eqnarray}
l_{a}l^{a} & = & n_{a}n^{a} = m_{a} m^{a} = 0, \; ~l_a n^a = 1, \nonumber \\
l_{a}m^{a} & = & n_{a}m^{a} = 0, \; m_{a} \bar{m}^{a} = -1, \nonumber \\
g_{ab} & = & l_{a}n_{b} + l_{b}n_{a} -  m_{a} \bar{m}_{b} - m_{b}
\bar{m}_{a},\nonumber \\
g^{ab} & = & l^{a}n^{b} + l^{b}n^{a} -  m^{a} \bar{m}^{b} - m^{b}
\bar{m}^{a}.
\end{eqnarray}
It turns out that the metric~(\ref{rknm}) satisfies the Einstein field equations
\begin{equation}\label{EEFE}
G_{ab} = T_{ab}^R+T_{ab}^{NR},
\end{equation}
where $T_{ab}^R= \chi(v,r,\theta) l_a l_b$ is the energy momentum
tensor of  null radiation, $\chi(v,r,\theta)$ the density of null
fluid and $T_{ab}^{NR}$ represents non-radiative field
\cite{Carmeli}.  In the stationary case the source, if it exists, is
the same for both a black hole and its rotating counterpart, e.g.,
vacuum for both Schwarzschild and Kerr black holes, and charge for
Reissner-Nordstr$\ddot{o}$m and Kerr-Newman black holes.  But, the
source for the Vaidya solution is just null radiation whereas it's
rotating counterpart (\ref{rknm}), in addition to  null radiation,
has a non-radiation field as source.  The  radiating Kerr black hole
metric (\ref{rknm}) is a natural generalization of the stationary
Kerr-black hole solutions \cite{kerr}, but it is Petrov type-II with
a twisting, shear free, null congruence the same as for Kerr black
hole, but the Kerr black hole is of Petrov type D. Further, all the
spin coefficients are identical to those for the Kerr black hole
\cite{Carmeli}. In addition, replacing $M(v)$ by  constant $M$ in
metric (\ref{rknm}), we get exactly the Kerr metric in original Kerr
coordinates. Further, it may be mentioned that the metric
(\ref{rknm}) is a radiating rotating metric, which also has the
correct static limit and it turns out that the stationary Kerr black
hole \cite{kerr} in
 Boyer-Lindquist coordinates $(t,\;r,\;\theta,\;\phi)$ can also be obtained by means of
local coordinate transformations and replacing $M(v)$ with constant
$M$ \cite{d'Inverno}. Further, in the limit $a=0$, the metric
(\ref{rknm}) reduces to the well known Vaidya metric.
\subsection{Rotating radiating Hayward black hole}
To avoid the black hole singularity problem, Hayward \cite{Hayward} proposed
 both static and radiating regular black hole models.  The radiating Hayward black hole solution  is given by the metric (\ref{Metric1}) with $f(v,r)$ defined by (\ref{fvr}) and $M(v)$ replaced by
\begin{equation}
M(v,r) = M(v) \frac{r^3}{r^3+q^3}.
\end{equation}
Here $M(v)$ is the radiating black hole mass and $q$ is a constant.
Next to get  radiating rotating regular black hole or rotating
radiating Hayward black hole \cite{Hayward}, we have to again start
with metric (\ref{Metric}), and when we apply the Newman-Janis
algorithm, as suggested by Newman and Janis \cite{nja}. For
generating radiating rotating
 regular black hole, following Bambi and Modesto
\cite{Bambi}, we must be able to recover the radiating Kerr black
hole or Carmeli solution (\ref{rknm}), in the limit $q=0$. Thus,
following this recipe, we again apply above complex transformation,
and as above demands the mass term $M(v,r)$ is invariant under the
transformation, and we get
\begin{equation}\label{nmass}
\mathcal{\bar{F}}(v,r,\theta) = 1- \frac{M(v,r) r }{\Sigma}.
\end{equation}
Hence the metric (\ref{rknm}) with the new mass function
(\ref{nmass}) is the  rotating radiating Hayward black hole, and in
the limit $q=0$, it goes over to the radiating Kerr black hole or
Carmeli's solution with mass $M(v)$. If the Einstein equations
 are used for the this radiating rotating regular black hole,
it is supported by stresses, e.g., the radial pressure $T^r_r$,
transverse pressure $T^{\theta}_{\theta}$, other stress such  as
$T^{\theta}_{\phi}$ etc. The supporting stresses (not all mentioned)
are given by
\begin{eqnarray*}
  T^r_r &=& \frac{6 M(v)g^3r^4}{\left(r^3+g^3\right)^3\Sigma^2}, \\
  T^{\theta}_{\theta} &=& \frac{6 M(v)g^3r^2 \left(a^2 \cos ^2 \theta (2 g^3-r^3)+r^2(g^3-2r^3)\right)}{\left(r^3+g^3\right)^3\Sigma^2},  \\
 T^{\theta}_{\phi} &=&  \frac{2 M(v)a^3r^4 \left({d M(v)}/{dv} \right)\cos \theta \sin^3 \theta}{\left(r^3+g^3\right)\Sigma^3}.
\end{eqnarray*}
These stresses fall off rapidly at large $r$ for $M(v),{d M(v)}/{dv}
\neq 0$.

\section{Physical parameters and horizons of Rotating Radiating Hayward black hole}
 Here we discuss the  physical properties of the metric of the radiating rotating regular black hole derived in the previous section. The easiest way to
detect a singularity, if it exists, in a spacetime is to observe the divergence of
certain invariants of the Riemann tensor.  We approach the singularity
problem by studying the behavior of the Ricci invariant $\mbox{R} = R_{ab}
R^{ab}$ ($R_{ab}$ is the Ricci tensor) and the Kretschmann invariant
$\mbox{K} = R_{abcd} R^{abcd}$ ($R_{abcd}$ is the Riemann tensor). For
the metric (\ref{rknm}) they read as
\begin{eqnarray}\label{invar}
& & \mbox{R} = 288 M^2(v) r^4 g^6 \frac{A \cos^4 \theta + B \cos^2 \theta +C }{(r+g)^6(r^2-rg+g)^6 \Sigma^4}, \nonumber \\
& & \mbox{K} = 48 M^2(v)r^4 \nonumber \\ && \times \frac{D \cos^8 \theta + E \cos^6 \theta +F \cos^4 \theta +G \cos^2 \theta + H}{(r+g)^6 \Sigma^6},
\label{eq:ks}
\end{eqnarray}
where $A \ldots H$ are functions of $r$ given by
\begin{eqnarray*}
  A &=& \left( {g}^{3}-\frac{{r}^{3}}{2} \right) ^{2} r^2,\; B=\left( -2\,{r}^{3}+{g}^{3} \right)  \left( {g}^{3}-\frac{{r}^{3} }{2}\right) {r}^{2}{a}^{2},
 \\
  C &=& \frac{{r}^{4} }{2}\left( -{r}^{3}{g}^{3}+{g}^{6}+\frac{{5r}^{6}}{2} \right),\; D=12\,{a}^{8}{g}^{6} \left( {g}^{3}-\frac{{r}^{3}}{2} \right) ^{2},    \\
  E &=& -4\,{a}^{6}{r}^{2} \left( {g}^{12}+\frac{{r}^{12}}{4}+{r}^{9}{g}^{3}+\frac{9{g}^{6}{r}^{6}}{4}+{\frac {59}{2}}
\,{r}^{3}{g}^{9} \right),  \\
  F&=&  22\,{a}^{4} \left(\frac{21{g}^{6}{r}^{6}}{2}+\frac{3{r}^{3}{g}^{9}}{11}+{\frac {15}{22}}\,{r}^{12}+{\frac {50}{11}}\,{r}^{9}{g}^{3}+{g}^{12}
 \right) {r}^{4},  \\
    G&=& 8\, \left( \frac{-7{r}^{9}{g}^{3}}{2}-{\frac {15}{8}}\,{r}^{12}+{\frac {69}{8}}
\,{g}^{6}{r}^{6}+{g}^{12}-\frac{9{r}^{3}{g}^{9}}{4} \right) {a}^{2}{r}^{6}, \\
H & = & 2\, \left( -{r}^{3}{g}^{9}+9\,{g}^{6}{r}^{6}+{g}^{12}-2\,{r}^{9}{g}^{3}+\frac{{r}^{12}}{2}
 \right) {r}^{8}.
\end{eqnarray*}
It is sufficient to study the Kretschmann and Ricci scalars for the
investigation of the spacetime curvature singularity(ies).
These invariants are regular everywhere including the origin $r=0$ for $a,\; M(v),\neq0$.
Further in the limit $\theta=\pi/2$ or $a=0$, they have the simple form
\begin{eqnarray}
  \mbox{R} &=& 72\,{\frac {{g}^{6} \left( M \left( v \right)  \right) ^{2} \left( -2
\,{r}^{3}{g}^{3}+5\,{r}^{6}+2\,{g}^{6} \right) }{ \left( r+g \right) ^
{6} \left( {r}^{2}-rg+{g}^{2} \right) ^{6}}},
 \\
  \mbox{K} &=&  48\,{\frac { \left( {r}^{12}-4\,{r}^{9}{g}^{3}+2\,{g}^{12}-2\,{r}^{3}{
g}^{9}+18\,{g}^{6}{r}^{6} \right)  \left( M \left( v \right)  \right)
^{2}}{ \left( {r}^{3}+{g}^{3} \right) ^{6}}}.
\end{eqnarray}
Thus the invariants are everywhere regular for $g\neq0$.  Further,
it is easy to obtain these invariants for the radiating Kerr BLACK
HOLE, in the limit $g=0$ in (\ref{invar}), and they read
\begin{eqnarray}
  \mbox{R} &=& 0, \\
  \mbox{K} &=& \frac{-48 M^2(v) }{\Sigma^{6}}\, H(r,\theta),
\end{eqnarray}
with $$ H(r,\theta)={ \left( {a}^{6} \cos^{6} \theta
-15\,{r}^{2}{a}^{4} \cos^{4} \theta+15\,{r}^{4}{a}^{2} \cos^{2}
\theta-{r}^{6} \right). }$$ Thus for the radiating Kerr black hole
$\Sigma=0$ happens to be a scalar polynomial singularity, and  such
a singularity is given by $r=0,\, \theta=\pi/2$. The set of points
given by $r = 0$ and $\theta=\pi/2$ represent a ring in the
equatorial plane of radius $a$ centered on the rotation axis of the
black hole, similar to what happens in the stationary Kerr black
hole \cite{ggsh}.

Inspired by the procedure in  Ref.~\cite{bc,jy}, a null vector
decomposition of the radiating regular Kerr metric (\ref{rknm}) is
of the form
\begin{equation}\label{gab}
g_{ab} = - n_a l_b - l_a n_b + \gamma_{ab},
\end{equation}
where $\gamma_{ab} = m_{a} \bar{m}_{b} + m_{b} \bar{m}_{a}$. Next we
calculate all physical parameters which in turn will help us to
study the horizon structure of a radiating Kerr-like regular black
hole. The optical behavior of null geodesic congruences is mastered
by the Raychaudhuri equation \cite{jy,rm,bdk,xd98,xd99}
\begin{equation}\label{re}
   \frac{d \Theta}{d v} = \kappa \Theta - R_{ab}l^al^b-\frac{1}{2}
   \Theta^2 - \sigma_{ab} \sigma^{ab} + \omega_{ab}\omega^{ab},
\end{equation}
with expansion $\Theta$, twist $\omega$, shear $\sigma$, and surface
gravity $\kappa$. In our discussion, the surface gravity \cite{jy}
is
\begin{equation}\label{sg}
\kappa = - n^a l^b \nabla_b l_a,
\end{equation}. The expansion \cite{jy} of the null rays,
parameterized by $v$, is given by
\begin{equation}\label{theta}
\Theta = \nabla_a l^a - \kappa,
\end{equation}
where $\nabla$ is the covariant derivative. The shear \cite{jy}
takes the form
\begin{equation}\label{shear}
\sigma_{ab} =  \Theta_{ab} - \Theta (\gamma_c^c) \gamma_{ab}.
\end{equation}
The luminosity due to loss of mass reads $L_M = - dM/dv$, $L_M < 1$
, which is measured in the  region where $d/dv$ is timelike
\cite{jy,rm,bdk}.

If we consider radiating regular black holes, it is useful to
discuss not only black hole solutions but their horizon structure.
In this section, we explore horizons of the radiating regular
Hayward black hole, and discuss the effects which comes from the
parameter $q$. In general, a  black hole has three important
surfaces \cite{jy}: timelike limit surface (TLS), apparent horizon
(AH) and event horizon (EH). For the non-radiating Schwarzschild
black hole, the three surfaces EH, AH, and TLS coincide. For the
Vaidya black hole which radiate, we have AH=TLS, but the EH is
different from AH. If we break spherical symmetry, but preserving
stationarity, e.g., Kerr black hole, then AH=EH but EH $ \neq $ TLS.

 Here we shall focus on the investigation of  these horizons for the radiating  regular Kerr-like black hole.
 As suggested by York \cite{jy}, three horizons may be obtained to
 $O(L)$ by noting that (i) for a radiating black hole,
 we can define  TLS as the locus where $g(\partial_v, \partial_v) = g_{vv} = 0$, AHs are termed as
surfaces such that $\Theta \simeq 0$, and EHs are surfaces such that
$d \Theta /dv \simeq 0$.

\begin{figure*}
\begin{center}
\begin{tabular}{c c c}
\includegraphics[scale=0.7]{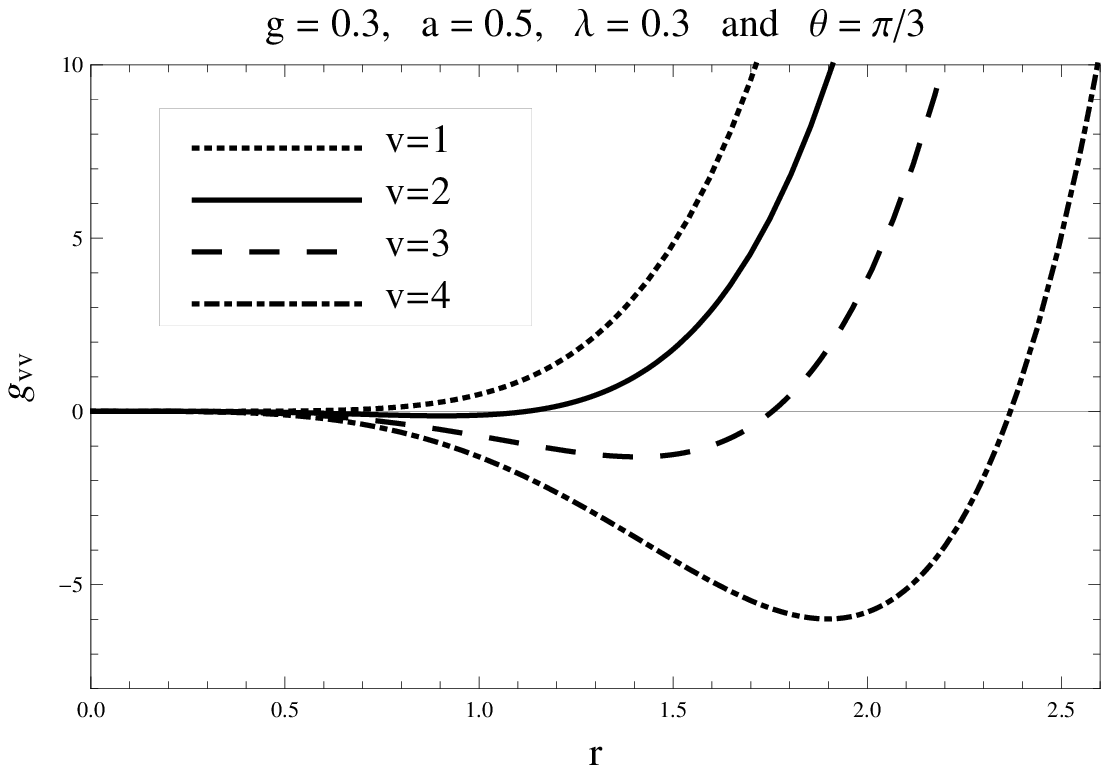}
&\includegraphics[scale=0.7]{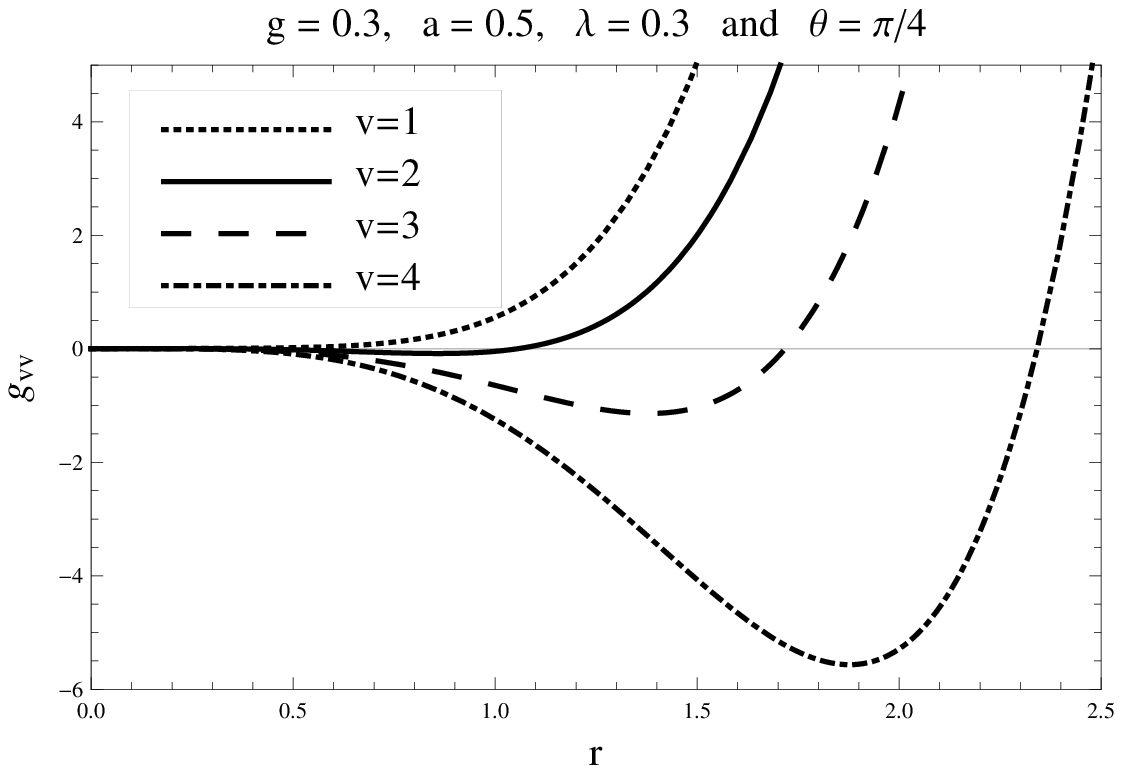}
 \\
 \includegraphics[scale=0.7]{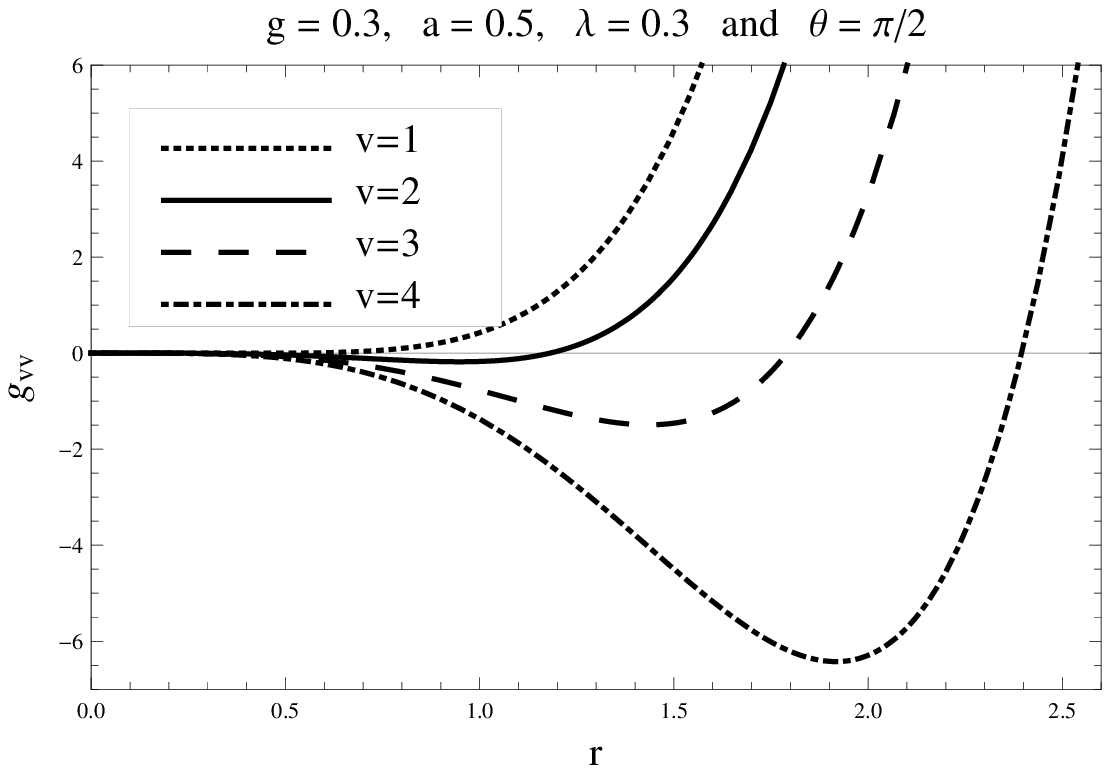}
 &\includegraphics[scale=0.7]{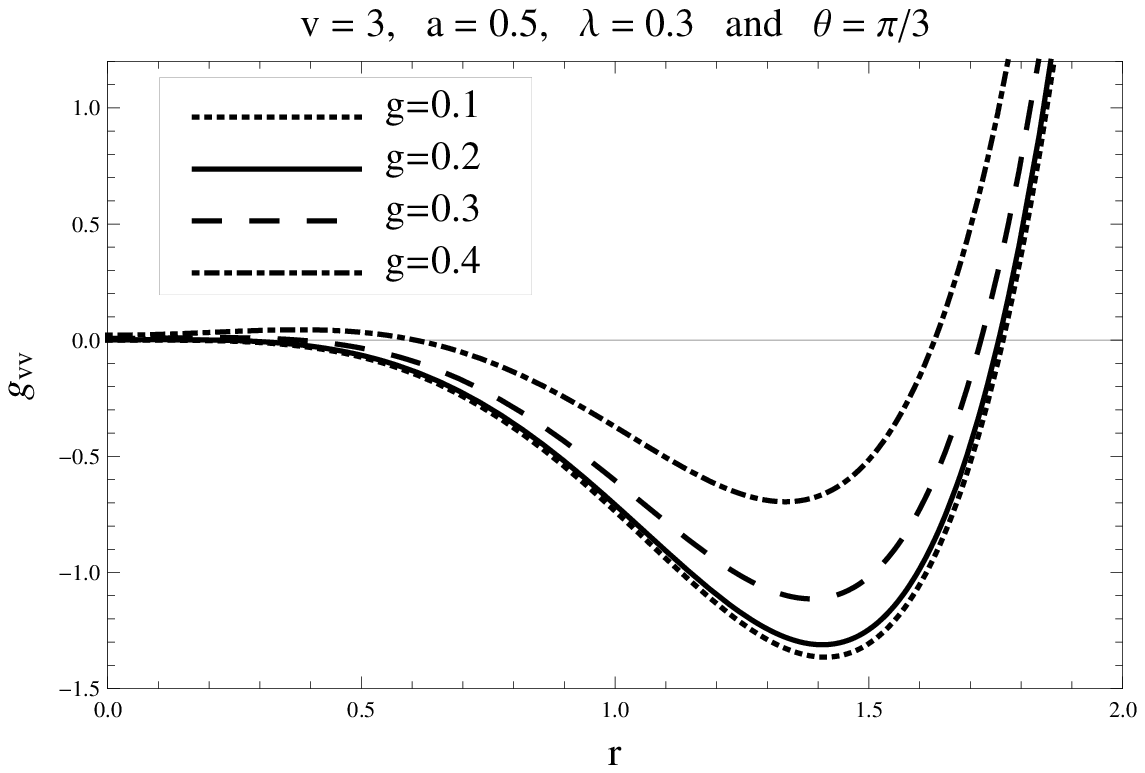}
  \\
  \includegraphics[scale=0.7]{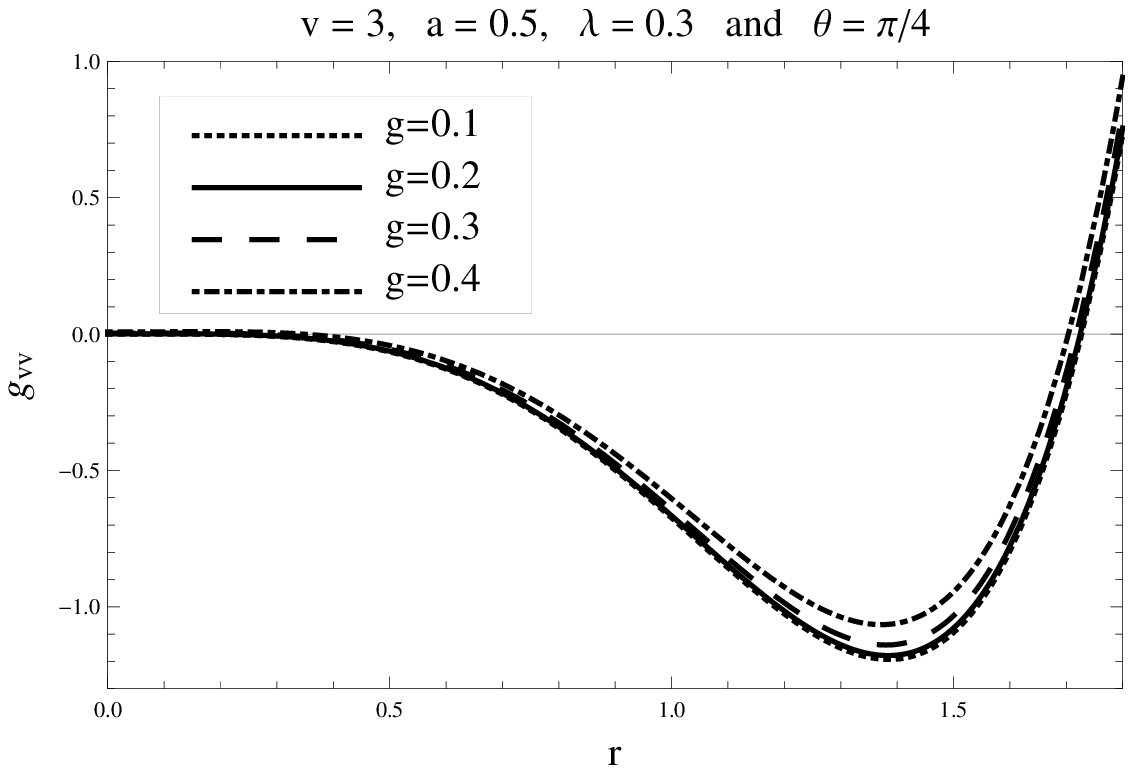}
  &\includegraphics[scale=0.7]{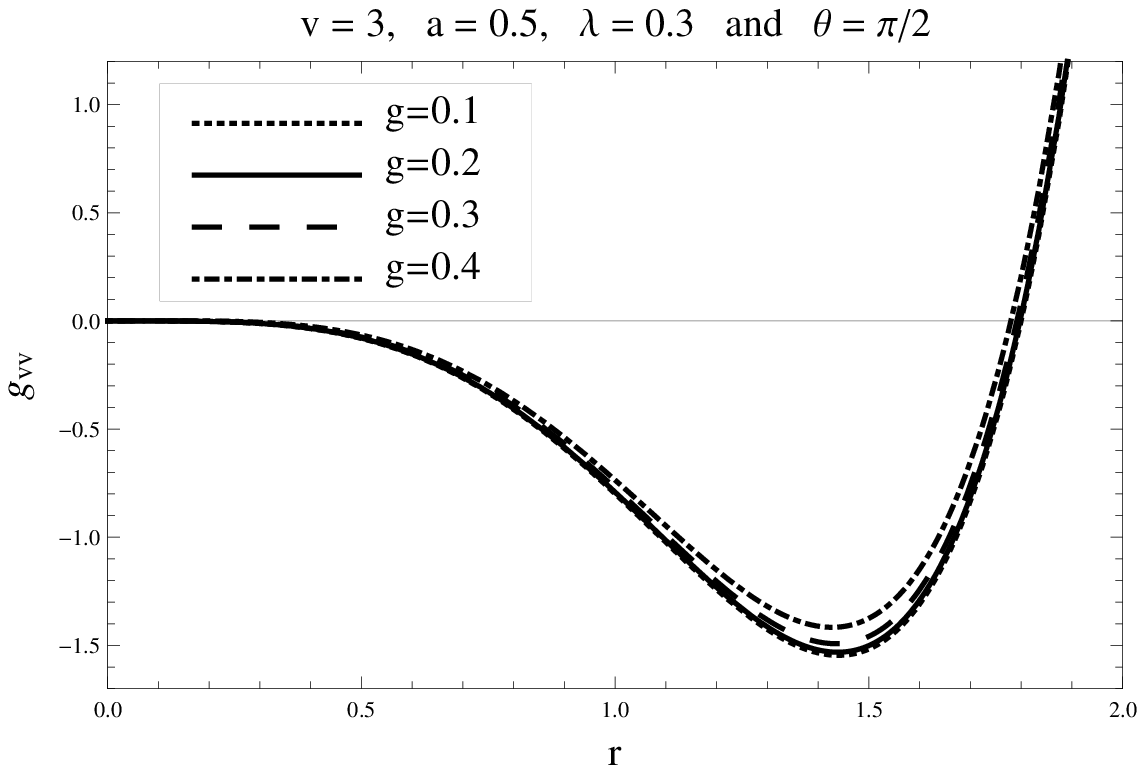}
   \\
   \includegraphics[scale=0.7]{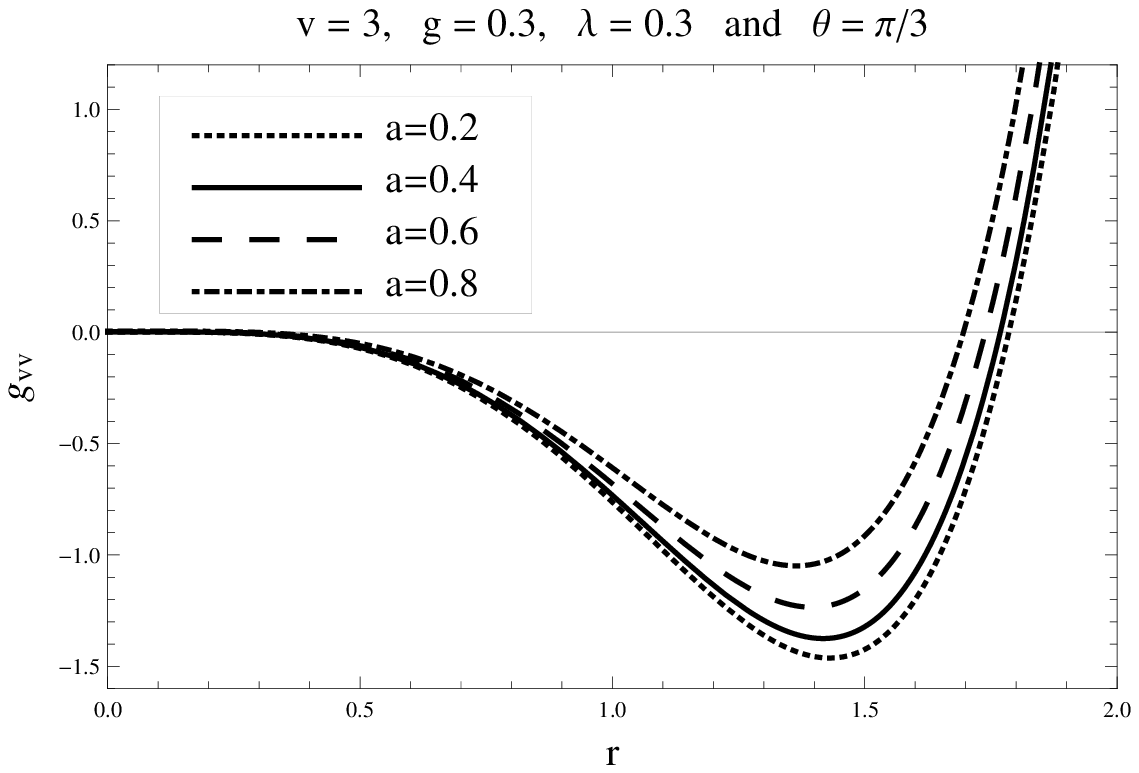}
   &\includegraphics[scale=0.7]{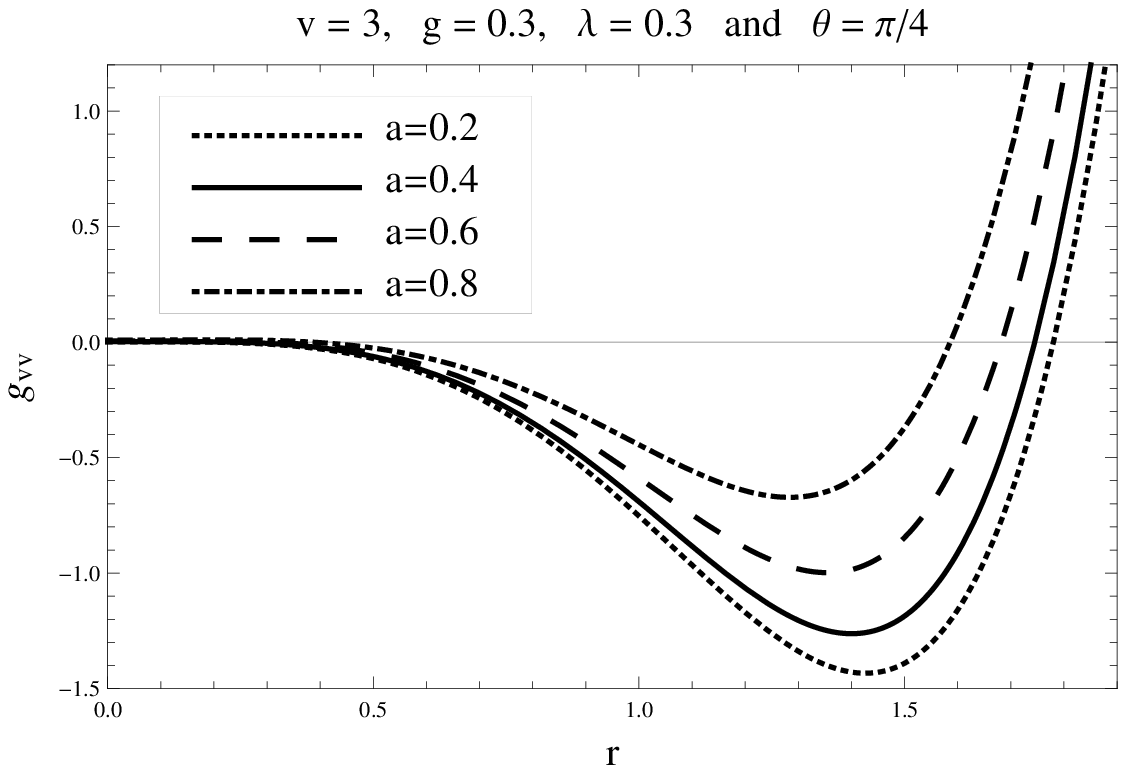}
 \end{tabular}
 \caption{Plots showing the timelike limit surface (TLS) ($g_{vv}$ vs $r$) for
 radiating rotating regular black hole  }\label{gvv_plot}
 \end{center}
\end{figure*}
%%%%%%

The TLS can be null,spacelike or timelike
 \cite{jy}. First, we find location of the TLS surface,
which for the radiating regular Kerr-like black hole requires that
the prefactor of the $dv^2$ term or $g_{vv}$ in the metric vanishes.
It follows from Eq.~(\ref{rknm}) that the TLS will satisfy $\Delta -
a^2 \sin^2 \theta  = 0$ \cite{xd99}, which can be written as
\begin{eqnarray}\label{tls1}
g_{vv} = & r^2 + a^2 \cos^2 \theta - 2 M(v,r) r   = 0.
\end{eqnarray}
On substituting $M(v,r)$ in Eq.~(\ref{tls1}), we produce a quintic equation of the form
\begin{equation}\label{tls2}
(r^3+g^3)(r^2+a^2 \cos^2 \theta)- 2 M(v) r^4=0.
\end{equation}
  It is not easy to solve Eq.~(\ref{tls1}) exactly, and hence we have solved it numerically and plotted the behaviors. In
 Fig.~\ref{gvv_plot}, for a given set of parameters, we show that two positive roots of
Eq.~(\ref{tls1}) are possible, i.e., the solution has two TLS, outer
and inner TLSs of radiating Kerr-like regular black hole. As
mentioned above, in the limit $g \rightarrow 0$, we get the
radiating Kerr-black hole solution \cite{sgsd}, and Eq.~(\ref{tls2})
takes the form
\begin{equation}\label{tlsnr}
 r^2 + a^2 \cos^2 \theta - 2 M(v)r =0.
\end{equation}
This is trivially solved to give
\begin{eqnarray} \label{tlskn}
  r_{TLS}^{-} &=& M (v) -\sqrt {  M^{2}(v)   -
 {a}^{2} \cos^{2} \theta },
 \nonumber \\
  r_{TLS}^{+} &=& M (v) + \sqrt {M^{2}(v)  -
 {a}^{2} \cos^{2} \theta}.
\end{eqnarray}
These are regular outer and inner TLSs for a radiating Kerr black
hole \cite{ggsh}. Further  in the non-rotating limit $a \rightarrow
0$, the solutions (\ref{tlskn}) reduce to
\begin{eqnarray} \label{tlskn1}
  r_{TLS}^{\pm} &=& 2 M (v),
\end{eqnarray}
which are TLSs of the Bonnor-Vaidya black hole.  Thus the radiating
regular Kerr-like black hole, in the GR limit and $a \rightarrow 0$,
reduces to the Vaidya black hole \cite{dg}. The TLSs of the
radiating regular Kerr-like  black hole is shown in
Fig.~\ref{gvv_plot} for different values of $g$ and rotation
parameter $a$, and it also shows the TLSs for
 variable time $v$. For definiteness we choose $M(v) \sim \lambda v +
 O(v)$.

 The AHs are defined as
surfaces such that $\Theta \simeq 0$ \cite{jy}.
 The AH can be either  space like or null, i.e., it can
'move' causally or acausally \cite{jy}.  The AH is the outermost
marginally trapped surface for the outgoing photons. Using
Eqs.~(\ref{nvector}) and (\ref{sg}), we get the surface gravity as
\begin{equation}\label{sge}
\kappa = \frac{1}{2\Sigma} \left[ \frac{\partial \Delta}{\partial
r} - \frac{2r}{\Sigma} \Delta\right],
\end{equation}
which on inserting the $\Delta$ expression takes the form
\begin{eqnarray}
\kappa &=& -{\frac {{a}^{2}r}{{\Sigma}^{2}}}+{\frac {- \left( {\frac {\partial }{
\partial r}}M \left( v,r \right)  \right) r-M \left( v,r \right) +r}{
\Sigma}} \nonumber \\ & & +{\frac {2\,M \left( v,r \right) {r}^{2}-{r}^{3}}{{\Sigma}^{2}.
}}
\end{eqnarray}
Eqs.~(\ref{nvector}), (\ref{theta}) and (\ref{sge}) then yield
\begin{eqnarray}\label{thetas}
\Theta &= & - \frac{r}{\Sigma^2} \Delta = -{\frac {r \left( {r}^{2}+{a}^{2}-2\,M \left( v,r \right) r \right) }{
{\Sigma}^{2}}}
\end{eqnarray}
It is obvious that the AHs are zeros of $\Theta=0$. Thus from
Eq.~(\ref{thetas}), the AH's are  zeros of
\begin{equation}\label{ahe}
(r^3+g^3)(r^2+a^2)- 2 M(v) r^4=0 = 0.
\end{equation}

\begin{figure*}
\begin{center}
\begin{tabular}{c c c}
\includegraphics[scale=0.7]{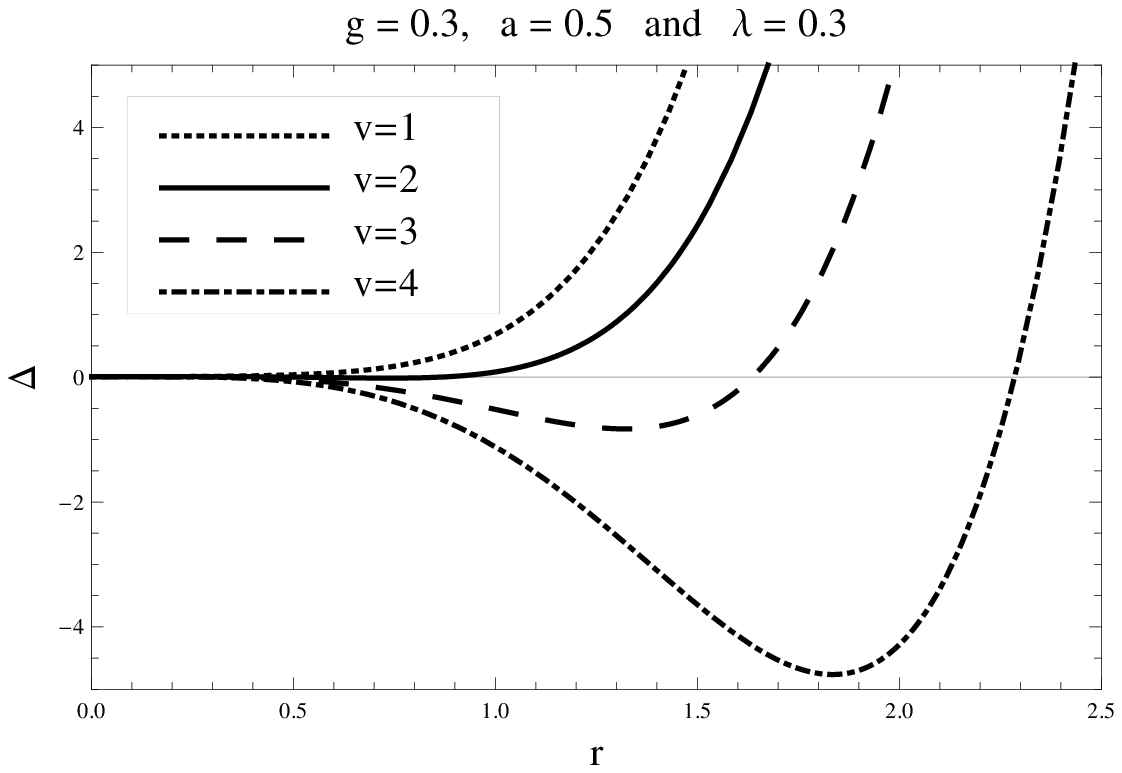}
&\includegraphics[scale=0.7]{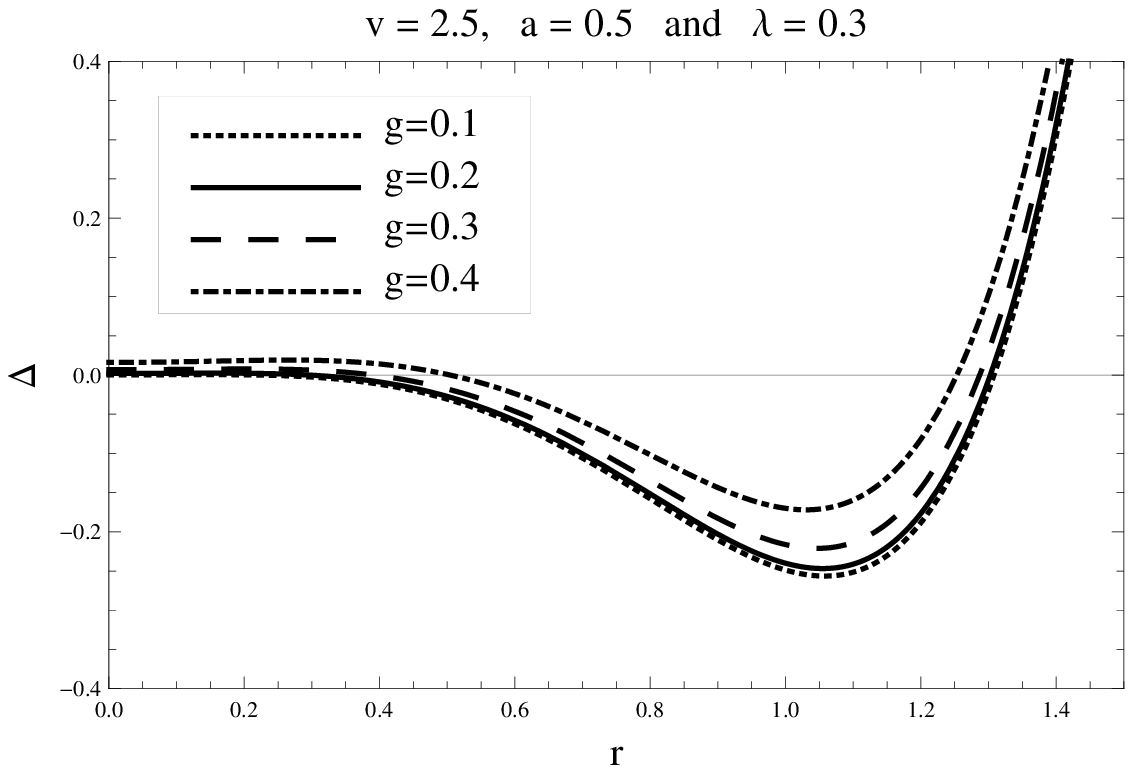}
 \\
 \includegraphics[scale=0.7]{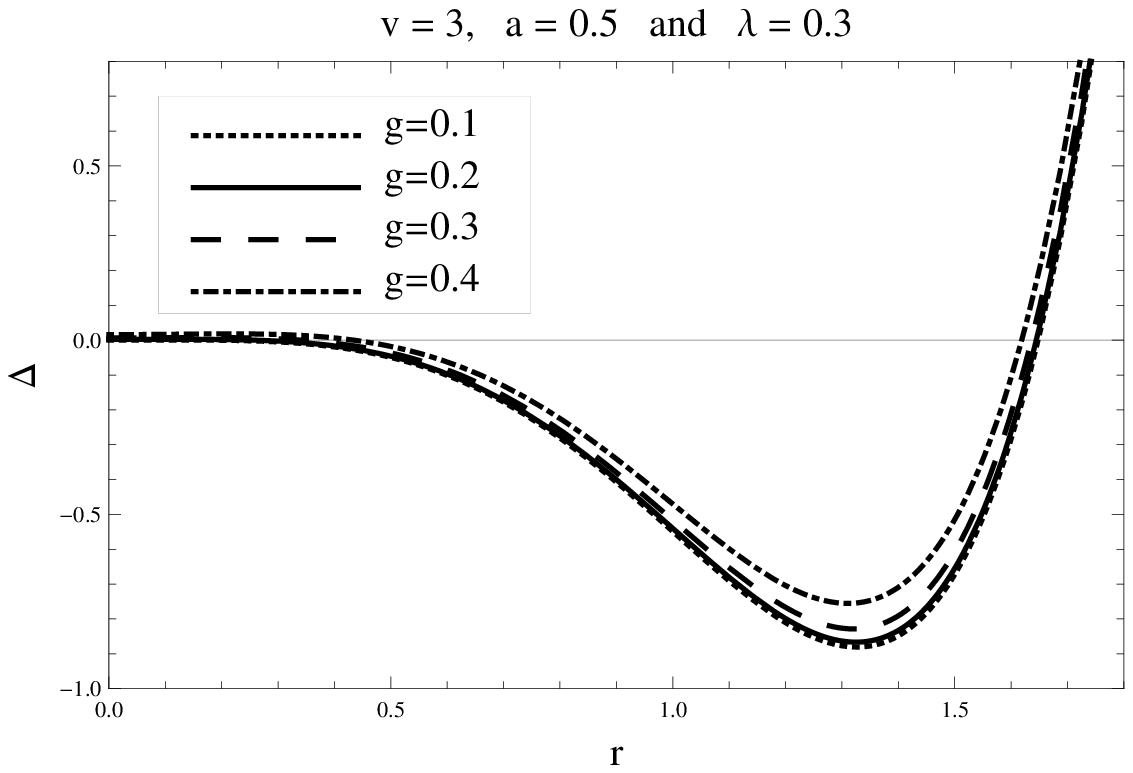}
 &\includegraphics[scale=0.7]{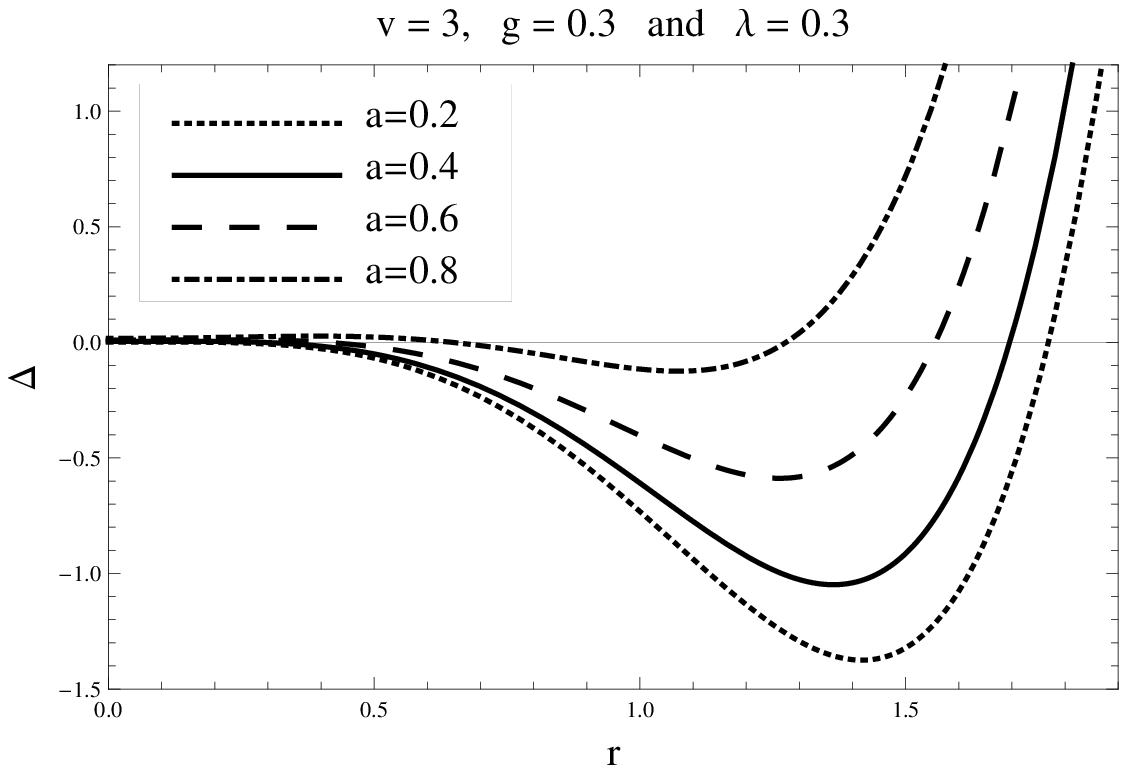}
 \end{tabular}
 \caption{Plots showing the AH ($\Delta$ vs $r$) for
 radiating rotating regular black hole }\label{AH}
 \end{center}
\end{figure*}
%%%%%%

Again in the limit $g \rightarrow 0$, we get
\begin{equation}\label{ahgr}
r^2- 2 M(v) r + a^2  = 0,
\end{equation}
which admits solutions
\begin{eqnarray} \label{ahkn}
  r_{AH}^{-} &=& M (v) -\sqrt {  M^{2}(v)   -
 {a}^{2} },
 \nonumber \\
  r_{AH}^{+} &=& M (v) + \sqrt {M^{2}(v)  -
 {a}^{2} }.
\end{eqnarray}
There exist sets of parameters for which two positive roots exist
 as shown in the Fig.~\ref{AH}. Unlike, TLS, the AH's are not $\theta$ dependent. Hence, contrary to
non-rotating radiating black holes, TLS, the AH  do not coincide in
the rotating radiating case.  The two roots correspond to inner and
outer AHs of black holes. The AH for radiating Kerr-like regular
black hole is depicted in Fig. \ref{AH} for different values of $g$
and rotation parameter $a$, and in the same Fig., we also show the
AH for different values of rotation parameter $a$ and time $v$. The
two roots corresponds to, respectively, outer and inner AHs for a
radiating Kerr-like regular black hole, and further in the
non-rotating limit $a \rightarrow 0$, the solutions (\ref{ahkn})
correspond to  AHs of black hole due to Vaidya. Further,
Eq.~(\ref{ahkn}) in the limit $a \rightarrow 0$ becomes exactly
Eq.~(\ref{tlskn}). Thus AHs coincide with TLSs, for the nonrotating
but radiating Vaidya case \cite{sgsd}.  In the stationary case $M$
is constant whereas in the radiating case $M(v)$ is a function of
the retarded time $v$.
\subsection{Event Horizon}
The above discussion, regarding TLSs and AHs,  is also true for the
stationary or non-radiating regular Kerr-like black solutions. The
AHs and EHs coincides for stationary black hole including regular
Kerr black hole. However, for the non-stationary or radiating Kerr
black hole, the three surfaces AH $\neq$ TLS $\neq$ EH and they are
susceptible to any kind of perturbations. Thus, Eqs.~(\ref{tls1})
and (\ref{ahe}) are the same as derived for the corresponding
stationary case  when $M(v)=M$ with $M$ constant. They are
determined via the Raychaudhuri Eq.~(\ref{re}) to $O(L)$. This
definition of the EH requires knowledge of the complete future of
the black hole. The EH is a null three-surface which is the locus of
outgoing future-directed null geodesic rays that never manage to
reach arbitrarily large distances from the black hole and behave
such that $d \theta / dv \simeq 0$. Hence, it's difficult to locate
the EH exactly in non-stationary spacetime. However, York \cite{jy}
gave a definition of the EH, which is in $O(L)$ equivalent to that
the photons at EH un-accelerated in the sense that
\begin{equation}\label{EH}
\frac{d^2r}{d n^2}_{|r=r_{EH}} \approx 0,
\end{equation}
with $d/dn = n^a \nabla_a$.  This criterion enables us to
differentiate  the AHs and the EHs to the necessary accuracy. It is
known that \cite{xd99}
\begin{equation}\label{rdd}
\frac{d^2r}{d n^2} = \frac{1}{\sqrt{A}2\Sigma^2} (r^2 + a^2)
\frac{\partial \Delta}{\partial v} + \frac{\Delta}{2 \Sigma}
\kappa.
\end{equation}
We note that for low luminosity the expression for the EH can be
obtained to $O(L)$ \cite{dg,xd98,xd99} after evaluating the surface
gravity $\kappa$ at the AH. Then the Eqs.~(\ref{rdd}), (\ref{sge}),
and the expression for $\Delta$ implies
\begin{equation}\label{ehe}
(r^3+g^3)(r^2+a^2)- 2 M^*(v) r^4=0,
\end{equation}
where
\begin{eqnarray*}
M^*(v) &=& M(v) + \frac{(r^2 + a^2)}{\sqrt{A}\; \kappa\; \Sigma} L.
\end{eqnarray*}
The EHs of the radiating regular Kerr-like black hole are zeros of
Eq.~(\ref{ehe}) which has interesting  mathematical similarity with
its counterpart Eq.~(\ref{ahe}) for AHs. But they are exactly same
for the stationary Kerr BH ($L=0$), but quite different for
radiating Kerr-like black holes. Also, in contrary to the AHs, EHs
have $\theta$ dependence as $M^*(v)$ involve $\Sigma$ or $\theta$. F
Thus the expression of the EH is exactly
 the same as its counterpart AH given by Eq.~(\ref{ahe}) with the mass
replaced by the effective mass $M^*(v)$ \cite{rm,xd99}. Thus, unlike
the stationary case, where AH=EH $\neq$ TLS, we have shown that for
radiating regular Kerr-like black hole, AH $\neq$ EH $\neq$ TLS. The
region bounded by the horizon and TLS is called the quantum
ergosphere.

\section{Conclusion}
The rotating Kerr black hole relish many useful properties distinct
from
 the non-rotating counterpart Schwarzschild black hole. However, there is a
surprising connection between the two different black holes of
general relativity, which analyzed by Newman and Janis \cite{nja} in
 their famous paper. They explicitly demonstrated that by applying a set of complex transformation,
it was possible to construct both the Kerr starting from the
Schwarzschild metric and likewise and Kerr-Newman solutions
beginning with Reissner-Nordstr$\ddot{o}$m metric\cite{nja}.

The Newman-Janis algorithm  is fruitful in deriving several rotating
black hole solutions starting from their non-rotating counterparts
\cite{Bambi,Toshmatov:2014nya,nja,sgsd}, which also includes the
rotating regular black hole \cite{Bambi,Toshmatov:2014nya}. The
algorithm is very useful since it directly allows us to generate
rotating black holes, which otherwise could be extremely tiresome
due to the nonlinearity of field equations. For  a review on the
Newman-Janis algorithm see, e.g., \cite{d'Inverno}.  In this paper,
we have generated a radiating (non-static) Kerr-like regular black
hole metric, which contains the radiating Kerr metric  as the
special case  when the deviation $g$ vanishes, and also the standard
Kerr metric when, in addition to $g=0$, the mass function $M(v)=M$
is constant.
 This metric does not arise from any particular set of field equations,
but the Newman-Janis algorithm works on the spherical radiating
solution to generate radiating rotating solutions. Thus, the derived
radiating Kerr-like regular metric (\ref{rknm}) bears the same
relation with the rotating regular black hole as does the Vaidya
metric to the Schwarzschild metric.

The structure of three surfaces TLSs, AHs, and EHs of the derived
radiating Kerr-like black hole were investigated by the method
developed by York \cite{jy} to $O(L)$ by a null vector decomposition
of the metric. The analysis presented  for determining the structure
of the horizons is applicable to the stationary rotating regular as
well, but  AHs coincide with EHs because stationary black holes do
not accrete, i.e., $L=0$. However,  the three surfaces do not
coincide with each other for radiating Kerr-like black holes.  For
each of TLS, AH and EH, there exist two surfaces corresponding to
the two positive roots $r^{-}$ and $r^{+}$, and they can be viewed,
respectively,  as inner and outer black hole horizons. Thus it means
that the presence of the term $q$ also,  we can find values of
parameters so that the two inner and outer horizons still exist as
in the case of radiating Kerr black hole.

To conclude,  the solutions presented here provide necessary grounds
to further study geometrical properties, causal structures and
thermodynamics of these black hole solutions, which will be subject
of a future project. Further generalization of such regular black
hole solution is an important direction and will be subject of our
forthcoming papers.

\acknowledgements We would like to thanks Pankaj Sheoran for his
help in plots.  SDM also acknowledges that this work is based upon
research supported by the South African Research Chair Initiative of
the Department of Science and Technology and the National Research
Foundation

\noindent
\end{document}